\documentclass[10pt]{paper}

\usepackage{amssymb}

\usepackage{epsfig}
\usepackage{amsmath}

\begin{document}



\title{Asymptotic and factorial expansions of Euler series truncation errors via exponential polynomials}
\author{Riccardo Borghi}
\institution{
Dipartimento di Elettronica Applicata\\
Universit\`a degli Studi ``Roma Tre", Rome, Italy\\
e-mail: borghi@uniroma3.it}



\maketitle


\begin{abstract}
A detailed analysis of the remainder obtained by truncating the Euler series up to the $n$th-order term is presented. 
In particular, by using an approach recently proposed by Weniger, asymptotic expansions of the remainder, both in inverse powers and in inverse rising factorials of $n$, are obtained.
It is found that the corresponding expanding coefficients are expressed, in closed form, in terms of exponential polynomials, well known in combinatorics, and in terms of associated Laguerre polynomials, respectively. A study of the divergence and/or of the convergence of the above expansions is also carried out for positive values of the Euler series
argument.
\end{abstract}






\section{Introduction}
\label{intro}

We consider the so-called Euler series (ES henceforth), defined as
\begin{equation}
\mathcal{E}(z)=
\displaystyle\sum_{m=0}^\infty\,(-1)^m\,z^m\,m!,
\label{Euler}
\end{equation}
where $z$ is a nonnegative number (possibly complex). The series in 
Eq.~(\ref{Euler}) has a zero convergence radius and gives a coded representation of the function
\begin{equation}
\mathcal{E}(z)=
\displaystyle\frac{\exp(1/z)}{z}\,
E_1\left(\displaystyle\frac 1z\right),
\label{Euler.0}
\end{equation}
where $E_1(x)$ denotes the exponential integral, defined by~\cite{abramowitz} 
%
\begin{equation}
E_1(x)=\displaystyle\int_x^\infty\,
\displaystyle\frac{\exp(-t)}{t}\,\mathrm{d}t,
\label{ExpIntegralE}
\end{equation}
for $|\arg x|<\pi$. The function in Eq.~(\ref{Euler.0}) is 
called the antilimit of the series in Eq.~(\ref{Euler}) 
and, according to Euler, can be viewed as the mathematical entity whose 
expansion gives rise to the divergent series.\footnote{
There is a nice quotation of a letter from Euler to Goldbach (1745)
in Ref.~\cite{calicetiArXiv07}, which reads:
\emph{``Summa cuiusque seriei est valor expressionis illius finitae, ex cuius evolutione illa series oritur,''} which may be translated from the Latin language 
as: ``The sum of any given series is the value of the specific finite expression whose expansion gave rise to that same series.''} 
Incidentally, the ES represents the paradigm for many  other factorially divergent asymptotic inverse power series occurring in special function theory, 
or arising from large-order perturbation expansions in theoretical physics.

The resummation of factorially diverging series like the ES can be successfully operated through the use of different strategies~\cite{calicetiArXiv07}. Among them, those based on nonlinear sequence transformations have proved, especially in recent times, to achieve the retrieving action in an effective way. 
The common feature of several types of sequence transformations is the following decomposition of the $n$th-order partial sum $s_n=\sum_{k=0}^n\,a_k$ of the starting series:
\begin{equation}
s_{n}= s+r_{n},
\label{Euler.1}
\end{equation}
where $s$ denotes the antilimit and $r_n=-\sum_{n+1}^\infty\,a_k$ the $n$th-order remainder. For the ES the decomposition 
in Eq.~(\ref{Euler.1}) can be straightforwardly derived by writing
\begin{equation}
\begin{array}{l}
s_n=\displaystyle\sum_{m=0}^n\,(-z)^m m!,
\end{array}
\label{Euler.1.1}
\end{equation}
and by expanding the factorial as $m!=\int_0^\infty\,\mathrm{d}t\,\exp(-t)t^m$,
thus obtaining 
\begin{equation}
\begin{array}{l}
s_n=
\displaystyle\int_0^\infty\,\mathrm{d}t\,\exp(-t)\,
\displaystyle\sum_{m=0}^n\,(-zt)^m=\\
\\
=
\displaystyle\int_0^\infty\,\mathrm{d}t\,
\displaystyle\frac{\exp(-t)}{1+zt}-
(-z)^{n+1}\,
\displaystyle\int_0^\infty\,\mathrm{d}t\,
\displaystyle\frac{\exp(-t)\,t^{n+1}}{1+zt},
\end{array}
\label{Euler.1.2}
\end{equation}
where use has been made of the explicit expression of the partial sum 
of the geometric series, i.e.,
\begin{equation}
\begin{array}{l}
\displaystyle\sum_{m=0}^n\,x^m=
\displaystyle\frac{1-x^{n+1}}{1-x}.
\end{array}
\label{Euler.1.2.1}
\end{equation}
On comparing Eq.~(\ref{Euler.1.2}) with Eqs.~(\ref{Euler.0}) and
(\ref{Euler.1}), it is  seen at once that
\begin{equation}
\begin{array}{l}
s=\mathcal{E}(z),\\
\\
r_n=-(-z)^{n+1}\,
\displaystyle\int_0^\infty\,\mathrm{d}t\,
\displaystyle\frac{\exp(-t)\,t^{n+1}}{1+zt}.
\end{array}
\label{Euler.1.3}
\end{equation}
The development of nonlinear sequence transformations is connected to 
the theory of converging factors~\cite{airey,miller-52,dingle}. According to it,
the remainder $r_n$ is expressed as the product between the first term of the
series not included in the partial sum, $a_{n+1}$, and a converging factor, say
$\varphi_n$, chosen in such a way that the relationship
\begin{equation}
\begin{array}{l}
s_n=s+a_{n+1}\,\varphi_n,
\end{array}
\label{Euler.1.3.1}
\end{equation}
is satisfied. In particular, from Eq.~(\ref{Euler.1.3}) it follows that,
for the ES, an integral representation of the $n$th-order converging factor
(or terminant) is (see Ref. \cite{dingle}, Ch. 21)
\begin{equation}
\begin{array}{l}
\varphi_n=-\displaystyle\frac 1{(n+1)!}\,
\displaystyle\int_0^\infty\,\mathrm{d}t\,
\displaystyle\frac{\exp(-t)\,t^{n+1}}{1+zt}.
\end{array}
\label{Euler.1.3.2}
\end{equation}

The aim of the present work is to find expansions, both in inverse powers and in inverse rising factorials of $n$, of the converging factor of the ES in Eq.~(\ref{Euler}) in such a way that the $n$th-order remainder in Eq.~(\ref{Euler.1.3}) can be expressed through the forms
\begin{equation}
r_n = a_{n+1}\,\displaystyle\sum_{k=0}^\infty\,
\displaystyle\frac{c_k}{(n+\alpha)^k},
\label{appC.0.1.1}
\end{equation}
and
\begin{equation}
r_n = a_{n+1}\,\displaystyle\sum_{k=0}^\infty\,
\displaystyle\frac{d_k}{(n+\alpha)_k},
\label{appC.0.1.1.0}
\end{equation}
where $\alpha$ denotes a positive parameter, $(\cdot)_k$ denotes the Pochhammer symbol, defined by
\begin{equation}
(a)_k=\displaystyle\frac{\Gamma(a+k)}{\Gamma(a)},
\label{pochhammer}
\end{equation}
and where the infinite sequences $\{c_k\}$ and $\{d_k\}$ are \emph{independent} of $n$. 
In particular, the  last prescription
revealed crucial for the Levin~\cite{levinIJCM-74}
and the $\delta$ (or Weniger)~\cite{wenigerPR-89} transformations,
which are strictly related to Eqs.~(\ref{appC.0.1.1}) and~(\ref{appC.0.1.1.0}), respectively, 
to be derived.\footnote{For an extensive review about applications of the Weniger transformation, see for instance Refs.~\cite{calicetiArXiv07,wenigerReviewJMP-04,borghiPRE-08,borghiPRE-09} and references therein.}
On the other hand, it should also be noted that different types of asymptotic expansions of the converging factors for the exponential integral function 
have already been found in the past~\cite{neuhaus-75,berg-77}, for instance by using the Airey's approach~\cite{airey}.

To find the closed-form expressions of the
sequences $\{c_k\}$ and $\{d_k\}$ we shall use the approach recently introduced by Weniger
in Ref.~\cite{weniger}.
According to it,  the starting point is the decomposition scheme in Eq.~(\ref{Euler.1}), 
from which it follows that the truncation error $r_n$ associated to the partial sum $s_{n}$ 
must satisfy the first-order difference equation
\begin{equation}
\begin{array}{l}
\Delta r_n=a_{n+1},
\end{array}
\label{appC.0.1}
\end{equation}
or
\begin{equation}
\begin{array}{l}
\displaystyle\frac{r_{n+1}-r_n}{a_{n+1}}=1,
\end{array}
\label{appC.0.1.0}
\end{equation}
where $\Delta$ denotes the forward difference operator with respect to $n$, 
i.e., such that $\Delta f(n)=f(n+1)-f(n)$.
In Ref.~\cite{weniger} such approach  was used to
reproduce the Euler-Maclaurin
formula for the remainder of the Dirichlet series for the Riemann zeta function. In the 
case of the ES, the same approach led to the exact expression of the first few terms
of expansions similar to those in Eqs.~(\ref{appC.0.1.1}) and~(\ref{appC.0.1.1.0}).
In the present work we will show that the whole sequences $\{c_k\}$ and 
$\{d_k\}$ appearing in Eqs.~(\ref{appC.0.1.1}) and~(\ref{appC.0.1.1.0}), respectively, can be obtained, for the case $\alpha=1$, through simple, analytical closed-form expressions. In particular, what we will find is that the $c_{k}$'s are expressed by exponential polynomials~\cite{bell,boyadzhievAAA-09,comtet}, while the coefficients $d_{k}$ 
turn out to be proportional to  associated Laguerre polynomials~\cite{abramowitz}. 

\section{Theoretical Analysis}
\label{wenigerEstimate}

\subsection{The Weniger approach for building up asymptotic 
expansions of truncation errors}
\label{wenigerApproach}

As anticipated in the previous section, the approach proposed by Weniger in Ref.~\cite{weniger} will be pursued to express, in closed-form terms, the 
two sequences $\{c_k\}$ and $\{d_k\}$ in Eqs.~(\ref{appC.0.1.1}) and~(\ref{appC.0.1.1.0}), for $\alpha=1$. We begin with the asymptotic inverse power series, needed for the Levin transformation. The inverse factorial expansion, related to the Weniger transformation, will be subsequently derived  starting from the former.
However, to obtain the asymptotic expansion of the converging factor as in Eq.~(\ref{appC.0.1.1}), following the prescriptions given in Ref.~\cite{weniger}, an intermediate step is necessary. It consists in replacing, in Eq.~(\ref{appC.0.1.0}), the remainder $r_n$ by the estimate, say $r^{(m)}_n$, given by
\begin{equation}
\begin{array}{l}
r^{(m)}_n \simeq a_n\,\displaystyle\sum_{k=0}^m\,
\displaystyle\frac{\gamma_k}{(n+1)^k}=
(-z)^n\,n!\,\displaystyle\sum_{k=0}^m\,
\displaystyle\frac{\gamma_k}{(n+1)^k},
\end{array}
\label{weniger.1}
\end{equation}
where $\{\gamma_0, \gamma_1, \ldots, \gamma_m\}$ are unknowns quantities 
which must be indipendent of $n$. 
It should be noted that
in Eq.~(\ref{weniger.1}) the truncation error of the ES is represented
as the last term \emph{included} in the partial sum multiplied by a truncated inverse power series in $n+1$. The correct converging factor will be then obtained starting from the knowledge of the $\gamma_k$'s.
The key point in the approach of Ref.~\cite{weniger}
consists in substituting from Eq.~(\ref{weniger.1}) into 
Eq.~(\ref{appC.0.1.0}) and in requiring that the subsequent equation be satisfied up to the power $n^{-m}$, i.e., that 
\begin{equation}
\begin{array}{l}
\displaystyle\frac{r^{(m)}_{n+1}-r^{(m)}_n}{a_{n+1}}=1+\mathcal{O}(n^{-m-1}),
\end{array}
\label{appC.0.1.0.1}
\end{equation}
for $n \to \infty$. After some algebra, it is obtained~\cite{weniger}
\begin{equation}
\begin{array}{l}
\displaystyle\frac{r^{(m)}_{n+1}-r^{(m)}_{n}}{(-1/x)^{n+1} (n+1)!}=
\displaystyle\frac{x}{n+1}\,\displaystyle\sum_{k=0}^m\,\displaystyle\frac{\gamma_k}{(n+1)^k}
+
\displaystyle\sum_{k=0}^m\,\displaystyle\frac{\gamma_k}{(n+2)^k}=
1+\mathcal{O}(n^{-m-1}),
\end{array}
\label{weniger.2}
\end{equation}
where, for convenience, it has been set $x=1/z$.
Note that Eq.~(\ref{weniger.2}) is asymptotically equivalent to
\begin{equation}
\begin{array}{l}
\displaystyle\frac{x}{n}\,\displaystyle\sum_{k=0}^m\,\displaystyle\frac{\gamma_k}{n^k}+\displaystyle\sum_{k=0}^m\,\displaystyle\frac{\gamma_k}{(n+1)^k}=
\displaystyle\sum_{k=0}^m\,\displaystyle\frac{\delta_{k,0}}{n^k}+\mathcal{O}(n^{-m-1}),
\end{array}
\label{weniger.3}
\end{equation}
which led to a linear system for the coefficients $\gamma_k$ that was explicitely solved by Weniger in Ref.~\cite{weniger} for $m=4$. 

We are now going to prove that such system 
can be solved in closed form for \emph{any} values of $m$. To show this, we start by writing
\begin{equation}
\begin{array}{l}
\displaystyle\frac{x}{n}\,\displaystyle\sum_{k=0}^m\,\displaystyle\frac{\gamma_k}{n^k}+\displaystyle\sum_{k=0}^m\,\displaystyle\frac{\gamma_k}{(n+1)^k}=
\gamma_0+\displaystyle\sum_{k=1}^{m+1}\,\displaystyle\frac{x\gamma_{k-1}}{n^k}+
\displaystyle\sum_{k=1}^{m}\,\displaystyle\frac{\gamma_{k}}{(n+1)^k},
\end{array}
\label{weniger.4}
\end{equation}
and we note that
\begin{equation}
\begin{array}{l}
\displaystyle\frac{1}{(n+1)^k}=
\displaystyle\frac{1}{n^k}\,
\displaystyle\frac{1}{(1+1/n)^k}=
\displaystyle\frac{1}{n^k}\,
\displaystyle\sum_{j=0}^\infty\,\displaystyle\frac{(-1)^j}{n^j}\,
\displaystyle\frac{(k)_j}{j!}=\\
 \\
 =
(-1)^k\,
\displaystyle\sum_{j=0}^\infty\,\displaystyle\frac{(-1)^{j+k}}{n^{j+k}}\,
\displaystyle\frac{(k+j-1)!}{j!(k-1)!}=
(-1)^k\,
\displaystyle\sum_{j=0}^\infty\,\displaystyle\frac{(-1)^{j+k}}{n^{j+k}}\,
\displaystyle\frac{(j+1)_{k-1}}{(k-1)!}
=\\
\\
=(-1)^k\,\displaystyle\sum_{j=k}^\infty\,
\displaystyle\frac{(-1)^j}{n^j}\,
\displaystyle\frac {(j-k+1)_{k-1}}{(k-1)!}=
(-1)^k\,\displaystyle\sum_{j=k}^\infty\,
\displaystyle\frac{(-1)^j}{n^j}\,
{\displaystyle{j-1}\choose{k-1}}.
\end{array}
\label{weniger.5}
\end{equation}
Furthermore, on substituting from Eq.~(\ref{weniger.5})
into Eq.~(\ref{weniger.4}), we have
\begin{equation}
\begin{array}{l}
\displaystyle\frac{x}{n}\,\displaystyle\sum_{k=0}^m\,\displaystyle\frac{\gamma_k}{n^k}+
\displaystyle\sum_{k=0}^m\,\displaystyle\frac{\gamma_k}{(n+1)^k}
=\\
\\
=\gamma_0+
\displaystyle\sum_{k=1}^{m+1}\,
\displaystyle\frac {x\,\gamma_{k-1}}{n^k}
+
\displaystyle\sum_{k=1}^{m}\,
\displaystyle\sum_{j=k}^\infty\,(-1)^{k+j}\,
\displaystyle\frac {\gamma_{k}}{n^j}
\displaystyle{{j-1}\choose{k-1}}=\\
\\
=\gamma_0+
\displaystyle\sum_{k=1}^{m}\,
\displaystyle\frac {x\,\gamma_{k-1}}{n^k}
+
\displaystyle\sum_{k=1}^{m}\,
\displaystyle\sum_{j=k}^m\,(-1)^{k+j}\,
\displaystyle\frac {\gamma_{k}}{n^j}
\displaystyle{{j-1}\choose{k-1}}\\
\\
+
\displaystyle\frac {x\,\gamma_{m}}{n^{m+1}}
+
\displaystyle\sum_{k=1}^{m}\,
\displaystyle\sum_{j=m+1}^\infty\,(-1)^{k+j}\,
\displaystyle\frac {\gamma_{k}}{n^j}
\displaystyle{{j-1}\choose{k-1}}
=\\
\\
=\gamma_0+
\displaystyle\sum_{k=1}^{m}\,
\displaystyle\frac {x\,\gamma_{k-1}}{n^k}
+
\displaystyle\sum_{k=1}^{m}\,
\displaystyle\sum_{j=k}^m\,(-1)^{k+j}\,
\displaystyle\frac {\gamma_{k}}{n^j}
\displaystyle{{j-1}\choose{k-1}}+
\mathcal{O}(n^{-m-1}),
\end{array}
\label{weniger.5.1}
\end{equation}
or, by interchanging the symbols $j$ and $k$,
\begin{equation}
\begin{array}{l}
\displaystyle\frac{x}{n}\,\displaystyle\sum_{k=0}^m\,\displaystyle\frac{\gamma_k}{n^k}+
\displaystyle\sum_{k=0}^m\,\displaystyle\frac{\gamma_k}{(n+1)^k}
=\\
\\
=\gamma_0+\displaystyle\sum_{k=1}^m\,
\displaystyle\frac 1{n^k}\,\left[x\,\gamma_{k-1}+
(-1)^k\,\displaystyle\sum_{j=1}^k\,(-1)^j\,
\displaystyle{{k-1}\choose{j-1}}
\gamma_j\right]
+
\mathcal{O}(n^{-m-1}).
\end{array}
\label{weniger.6}
\end{equation}
Equation~(\ref{weniger.6}), together with Eq.~(\ref{weniger.3}), leads to
the following linear system for the $\gamma_k$'s:
\begin{equation}
\begin{array}{l}
\gamma_0=1,\\
\\
x\,\gamma_{k-1}+(-1)^k\,
\displaystyle\sum_{j=1}^k\,(-1)^j\,
\displaystyle{{k-1}\choose{j-1}}
\,\gamma_j=0.
\end{array}
\label{weniger.7}
\end{equation}
The solution of such system can be expressed in closed form
simply by evaluating the quantity
\begin{equation}
\begin{array}{l}
-x\,\displaystyle\sum_{k=1}^n\,
\displaystyle{{n-1}\choose{k-1}}
\,\gamma_{k-1},
\end{array}
\label{weniger.7.1}
\end{equation}
which, by taking Eq.~(\ref{weniger.7}) into account,
takes on the form
\begin{equation}
\begin{array}{l}
-x\,\displaystyle\sum_{k=1}^n\,
\displaystyle{{n-1}\choose{k-1}}
\,\gamma_{k-1}=
\displaystyle\sum_{k=1}^n\,
\displaystyle\sum_{j=1}^k\,
(-1)^{k+j}\,
\displaystyle{{n-1}\choose{k-1}}\,
\displaystyle{{k-1}\choose{j-1}}\,\gamma_j=\\
\\
=
\displaystyle\sum_{j=1}^n\,
(-1)^j\,\gamma_j\,
\displaystyle\sum_{k=j}^n\,
(-1)^{k}\,
\displaystyle{{n-1}\choose{k-1}}\,
\displaystyle{{k-1}\choose{j-1}}=
\displaystyle\sum_{j=1}^n\,(-1)^{2j}\,\gamma_j\,\delta_{n,j}=
\gamma_n,
\end{array}
\label{weniger.9}
\end{equation}
where formula 4.2.4.45 of~\cite{PrudnikovI} has been used.
It is worth exploring Eq.~(\ref{weniger.9}) in a deeper way.
Actually, this equation already contains the closed-form expression
of the $\gamma_n$ coefficients, as we shall see in a moment.

\subsection{The asymptotic expansion of the truncation error and the 
exponential polynomials}
\label{asymptoticExpansion}

Due to its importance, we rewrite Eq.~(\ref{weniger.9}) as
\begin{equation}
\begin{array}{l}
\gamma_n=
-x\,\displaystyle\sum_{k=1}^n\,
\left({n-1}\atop{k-1}\right)\,\gamma_{k-1}.
\end{array}
\label{weniger.9.1}
\end{equation}
It must be noted that there exist a whole class of functions
satisfying the relation in Eq.~(\ref{weniger.9.1}). 
Such functions are called exponential, or Bell, polynomials~\cite{bell}, 
and will be denoted $\phi_n(x)$.%
\footnote{
Here and in the following we are going to use, for the Bell 
polynomials, the notation given in the recent review by Boyadzhiev~\cite{boyadzhievAAA-09}.
Furthermore, note that the numerical evaluation of Bell polynomials $\phi_n(x)$ of arbitrary order is currently implemented,
up to arbitrary precision, whithin the \emph{Mathematica} platform
through the command {\tt BellB[n,x]}.
}
These polynomials are defined through the following generating function formula:
\begin{equation}
\displaystyle\sum_{n=0}^\infty\,
\phi_n(x) \displaystyle\frac{t^n}{n!}=
\exp\left\{x\left[\exp(t)-1\right]\right\},
\label{generating.1}
\end{equation}
and have the explicit expansion
\begin{equation}
\phi_n(x) =
\displaystyle\sum_{k=0}^n\,
{S}(n,k)\,x^k,
\label{explicit}
\end{equation}
where 
${S}(n,k)$ denotes the Stirling number of the second kind, which is defined through\footnote{Here and in the rest of the paper we use, for the Stirling numbers of the first and of the second kind, the notation adopted in 
Ref.~\cite{comtet}.}
\begin{equation}
z^n= (-1)^n\,\displaystyle\sum_{k=0}^n\,(-1)^k\,
S(n,k)\,(z)_k.
\label{S2gen}
\end{equation}
Exponential polynomials satisfy the recurrence relationship
\begin{equation}
\phi_{n+1}(x) = x [\phi'_n(x)+\phi_n(x)],
\label{recurrenceBell}
\end{equation}
with $\phi_0(x)=1$ and, more importantly, they fulfill 
the relation
\begin{equation}
\phi_{n}(x) = 
x\,\displaystyle\sum_{k=1}^n\,
\displaystyle{{n-1}\choose{k-1}}\,\phi_{k-1}(x),
\label{recurrenceBell.2}
\end{equation}
which, when compared to Eq.~(\ref{weniger.9.1}),
shows that the coefficient $\gamma_k$ 
is proportional to  $\phi_k(-x)$ and, by virtue of the initial condition $\gamma_0=1$, that
\begin{equation}
\begin{array}{l}
\gamma_k=\phi_k(-x).
\end{array}
\label{weniger.10}
\end{equation}
Equation~(\ref{weniger.10}) represents one of the main results of the present work. According to it, 
we shall express the $n$th-order remainder of the ES  through 
the following asymptotic series:
\begin{equation}
r_n =
a_n\,\displaystyle\sum_{k=0}^\infty\,
\displaystyle\frac{\phi_k(-1/z)}{(n+1)^{k}},
\label{weniger.1Ter}
\end{equation}
which, as pointed out at the beginning of Sec.~\ref{wenigerApproach},
 is not yet of the 
desired form given in Eq.~(\ref{appC.0.1.1}). To find the correct
sequence $\{c_k\}$  for $\alpha=1$, it is sufficient to recast 
Eq.~(\ref{weniger.1Ter}) as
\begin{equation}
\begin{array}{l}
    r_n =-\displaystyle\frac{a_{n+1}}{z(n+1)} \,\displaystyle\sum_{k=0}^\infty\,
\displaystyle\frac{\gamma_{k}}{(n+1)^k}=
a_{n+1}\,\displaystyle\sum_{k=1}^\infty\,
\displaystyle\frac{-\phi_{k-1}(-1/z)/z}{(n+1)^k},
\end{array}
\label{WT.3}
\end{equation}
which, once compared to Eq.~(\ref{appC.0.1.1}), leads to
\begin{equation}
c_{k}=
\left\{
\begin{array}{lr}
0,&k=0,\\
&\\
-\displaystyle\frac{\phi_{k-1}(-1/z)}{z},&k>0.
\end{array}
\right.
\label{WT.3.1}
\end{equation}

\subsection{The factorial expansion}
\label{factorialExpansion}

The transformation of an inverse power series to a factorial series can be accomplished with the help of the Stirling numbers of the \emph{first} kind, say $s(n,j)$, which are defined 
through
\begin{equation}
\begin{array}{lcl}
(-1)^n\,(z)_n=
\displaystyle\sum_{j=0}^{n}\,
s(n,j)\,(-z)^j.
\end{array}
\label{weniger.1Bis.1}
\end{equation}
In particular, the Stirling numbers of the first kind occur in the factorial series 
expansion of an inverse power, namely~\cite{weniger}
\begin{equation}
\begin{array}{lcl}
\displaystyle\frac 1{\zeta^{k}}&=&
(-1)^k\,\displaystyle\sum_{j=k}^{\infty}\,
(-1)^j\,\displaystyle\frac{
s(j-1,k-1)}{(\zeta)_j}=
(-1)^k\,\displaystyle\sum_{j=1}^{\infty}\,
(-1)^j\,\displaystyle\frac{
s(j-1,k-1)}{(\zeta)_j},
\end{array}
\label{weniger.1Bis.1.0}
\end{equation}
valid for $k\ge 1$, where in the last passage use has been made of the fact that 
$s(n,m)=0$ when $m>n$. Accordingly, as pointed out in Ref.~\cite{weniger}, given an asymptotic power series of the form $\sum_{k=0}^{\infty}\,c_{k}/\zeta^{k}$, the following identity can be established:
\begin{equation}
\begin{array}{l}
    \displaystyle\sum_{k=0}^{\infty}\,\displaystyle\frac{c_{k}}{\zeta^{k}}=   \displaystyle\sum_{k=0}^{\infty}\,\displaystyle\frac{d_{k}}{(\zeta)_{k}},
\end{array}
\label{weniger.1Bis.1.1}
\end{equation}
where 
\begin{equation}
    d_{k}=\left\{
    \begin{array}{lr}
	c_{0},&k=0,\\
	&\\
	(-1)^{k}\,
	\displaystyle\sum_{j=1}^{k}\,
    (-1)^{j}\,
    s(k-1,j-1)\,c_{j},&k\ge 1.
    \end{array}
    \right.
\label{weniger.1Bis.2}
\end{equation}
Then, on substituting from Eq.~(\ref{WT.3.1}) into 
Eq.~(\ref{weniger.1Bis.2}), after some algebra it is found 
that the expanding coefficients $\{d_{k}\}$ in Eq.~(\ref{appC.0.1.1.0}) 
are given by
\begin{equation}
d_{k}=
\left\{
\begin{array}{lr}
0,&k=0,\\
&\\
-\displaystyle\frac{\psi_{k-1}(-1/z)}{z},&k>0,
\end{array}
\right.
\label{WT.5}
\end{equation}
where the function $\psi_k(x)$ is defined as
\begin{equation}
\psi_{k}(x)=
	(-1)^{k}\,
	\displaystyle\sum_{j=0}^{k}\,(-1)^{j}\,
	s(k,j)\,\phi_{j}(x).
\label{GR.2.15}
\end{equation}
On substituting from Eq.~(\ref{WT.5}) into Eq.~(\ref{appC.0.1.1.0}),
we eventually obtain
\begin{equation}
\begin{array}{l}
    r_n =a_{n+1}\,\displaystyle\sum_{k=1}^\infty\,
\displaystyle\frac{-\psi_{k-1}(-1/z)/z}{(n+1)_k}.
\end{array}
\label{WT.6}
\end{equation}
In the next section it will be proved that $\psi_n(x)$ 
is proportional to the associated Laguerre polynomial of orders $n$ 
and -1~\cite{abramowitz}.

\section{The $\psi_n(x)$ polynomials}
\label{properties}

From the definition given in Eq.~(\ref{GR.2.15}), on using Eq.~(\ref{explicit})
we have
\begin{equation}
\begin{array}{l}
\psi_{n}(x)=
	(-1)^{n}\,
	\displaystyle\sum_{j=0}^{n}\,(-1)^{j}\,
	s(n,j)\,
	\displaystyle\sum_{k=0}^j\,
	S(j,k)\,x^k=\\
	\\
	=
	(-1)^{n}\,
	\displaystyle\sum_{k=0}^n\,
	x^k\,
	\displaystyle\sum_{j=k}^{n}\,(-1)^{j}\,
	s(n,j)\,S(j,k)	=\\
	\\
	=
	(-1)^{n}\,
	\displaystyle\sum_{k=0}^n\,
	x^k\,
	\displaystyle\sum_{j=0}^{n}\,(-1)^{j}\,
	s(n,j)\,S(j,k),
	\end{array}
	\label{polExp.1}
\end{equation}
where in the last passage use has been made of the fact that 
${S}(j,k)=0$ for $k>j$. 
The expanding coefficients in Eq.~(\ref{polExp.1}) can be given a closed form. 
To show this, we first recall formula 24.1.4.C of Ref.~\cite{abramowitz}, i.e.,
\begin{equation}
\begin{array}{l}
	S(j,k)=
	\displaystyle\frac{(-1)^k}{k!}\,\displaystyle\sum_{\ell=0}^k\,
	(-1)^\ell\,\left(\displaystyle{k\atop\ell}\right)\,\ell^j,
	\end{array}
	\label{newFormulaStirling.1}
\end{equation}
which gives
\begin{equation}
\begin{array}{l}
	\displaystyle\sum_{j=0}^{n}\,(-1)^{j}\,
	s(n,j)\,{S}(j,k) =
	\displaystyle\frac{(-1)^k}{k!}\,
	\displaystyle\sum_{\ell=0}^k\,
	(-1)^\ell\,\left(\displaystyle{k\atop\ell}\right)\,
	\displaystyle\sum_{j=0}^{n}\,(-\ell)^{j}\,
	s(n,j)=\\
	\\
	=
	\displaystyle\frac{(-1)^{k+n}}{k!}\,
	\displaystyle\sum_{\ell=0}^k\,
	(-1)^\ell\,\left(\displaystyle{k\atop\ell}\right)\,
	(\ell)_n,
	\end{array}
	\label{newFormulaStirling.2}
\end{equation}
where in the last passage the definition of the generating function of the Stirling numbers of the first kind, given in Eq.~(\ref{weniger.1Bis.1}), has been used. Furthermore, on taking into account that
\begin{equation}
\begin{array}{l}
	(\ell)_n=n!\,\left(\displaystyle{{\ell+n-1}\atop n}\right),
	\end{array}
	\label{newFormulaStirling.3}
\end{equation}
Eq.~(\ref{newFormulaStirling.2}) becomes
\begin{equation}
\begin{array}{l}
	\displaystyle\sum_{j=0}^{n}\,(-1)^{j}\,
  s(n,j)\,S(j,k)=
	(-1)^{k+n}\,\displaystyle\frac{n!}{k!}\,
	\displaystyle\sum_{\ell=0}^k\,
	(-1)^\ell\,
	\left(\displaystyle{k\atop\ell}\right)\,
	\left(\displaystyle{{\ell+n-1}\atop n}\right),
	\end{array}
	\label{newFormulaStirling.4}
\end{equation}
and, by using formula 4.2.5.26 of Ref.~\cite{PrudnikovI}, it is obtained
\begin{equation}
\begin{array}{l}
	\displaystyle\sum_{j=0}^{n}\,(-1)^{j}\,
	s(n,j)\,S(j,k)=
	(-1)^n\,
	\left(\displaystyle{{n-1}\atop{n-k}}\right)\,\displaystyle\frac{n!}{k!}=(-1)^n\,
	\displaystyle{n\choose k}\,\displaystyle\frac{(n-1)!}{(k-1)!}.
	\end{array}
	\label{newFormulaStirling.5}
\end{equation}
On substituting from Eq.~(\ref{newFormulaStirling.5})
into Eq.~(\ref{polExp.1}), the polynomial expansion of $\psi_n(x)$ turns out to be, for $n>0$,
\begin{equation}
\begin{array}{l}
	\psi_n(x)=
	(n-1)!\,
	\displaystyle\sum_{k=1}^{n}\,
	\displaystyle\frac{x^k}{(k-1)!}\,\displaystyle{n\choose k},
	\end{array}
	\label{newFormulaStirling.6}
\end{equation}
from which it follows that
\begin{equation}
\psi_n(x)=n!\,L^{(-1)}_n(-x),
\label{generating.4.7}
\end{equation}
where $L^{(\alpha)}_n(\cdot)$ denotes the associated Laguerre polynomial of 
orders $n$ and $\alpha$~\cite{abramowitz}.
Finally, on substituting from Eq.~(\ref{generating.4.7}) into Eq.~(\ref{WT.6}),
the factorial expansion of the $n$th-order remainder of the ES reads
\begin{equation}
\begin{array}{l}
    r_n =-a_{n+1}\,\displaystyle\sum_{k=0}^\infty\,
\displaystyle\frac{k!}{(n+1)_{k+1}}\,L^{(-1)}_{k}(1/z)/z,
\end{array}
\label{generating.5}
\end{equation}
which, together with Eq.~(\ref{WT.3}), constitutes the main result of
the present paper. 

\section{Discussions}
\label{discussions}

\subsection{Preliminaries}
\label{prediscussions}

First of all, we note that the remainder $r_n$ in Eq.~(\ref{Euler.1.3}) can be evaluated in closed
form, and turns out to be given by
\begin{equation}
r_n=-a_{n+1}\,\displaystyle\frac{\exp\left(\displaystyle\frac 1z\right)}z\,E_{n+2}\left(\displaystyle\frac 1z\right),
\label{convergence.0}
\end{equation}
which, once compared with Eqs.~(\ref{WT.3})
and~(\ref{generating.5}) gives at once the following two 
expansions for the exponential integral function:
\begin{equation}
\begin{array}{l}
E_{n+1}(x)=\displaystyle\sum_{k=0}^\infty\,
\displaystyle\frac{\phi_k(-x)\,\exp(-x)}{n^{k+1}},
\end{array}
\label{convergence.1}
\end{equation}
and
\begin{equation}
\begin{array}{l}
E_{n+1}(x)=\displaystyle\sum_{k=0}^\infty\,
\displaystyle\frac{k!}{(n)_{k+1}}\,L^{(-1)}_k(x)\,\exp(-x),
\end{array}
\label{convergence.2}
\end{equation}
respectively.
Moreover, Eqs.~(\ref{convergence.1})
and~(\ref{convergence.2}) can be cast, by using the connection
between the exponential integral and the incomplete gamma functions~\cite{abramowitz},
in the following form:
\begin{equation}
\begin{array}{l}
\Gamma(-n,x)=x^{-n}\,\exp(-x)\,
\displaystyle\sum_{k=0}^\infty\,
\displaystyle\frac{1}{n^{k+1}}\,\phi_k(-x),
\end{array}
\label{convergence.1.bis}
\end{equation}
and
\begin{equation}
\begin{array}{l}
\Gamma(-n,x)=x^{-n}\,\exp(-x)\,
\displaystyle\sum_{k=0}^\infty\,
\displaystyle\frac{k!}{(n)_{k+1}}\,
L^{(-1)}_k(x),
\end{array}
\label{convergence.2.bis}
\end{equation}
respectively. In particular, the expansion in Eq.~(\ref{convergence.1.bis}) 
displays a structure very similar to an asymptotic expansion recently found in Ref.~\cite{boyadzhievJMMS-05} for the (lower) incomplete gamma function $\gamma(\lambda,x)$~\cite{abramowitz}, with $\lambda>0$.

\subsection{Analysis of the convergence of the two series for positive $x$}
\label{convergence}

It is worth studying the character of the series in Eqs.~(\ref{convergence.1})
and~(\ref{convergence.2}), for $x>0$.
As far as the series in Eq.~(\ref{convergence.1}) is concerned, 
we use the asymptotics, for $k \gg 1$, of the Bell polynomials, 
recently reviewed in Ref.~\cite{dominiciJCAM-09}. In particular, 
for $x>0$ we have~\cite{dominiciJCAM-09}
\begin{equation}
\begin{array}{l}
\exp(-x)\,\phi_k(-x) \approx k!\sqrt{\displaystyle\frac 2{\pi k}}\,\,
\displaystyle\frac{\exp\left\{k\left[\log\displaystyle\frac{\sin\varphi}{\varphi}
-\displaystyle\frac{\sin\varphi}{\varphi}\cos\varphi\right]\right\}}{\left[\left(\displaystyle\frac \varphi{\sin\varphi}-\cos\varphi\right)^2+\sin^2\varphi\right]^{1/4}}\,\\
\\
\times
\sin\left[k\left(\pi-\varphi+\displaystyle\frac{\sin^2\varphi}{\varphi}\right)+\eta(\varphi)\right],
\end{array}
\label{convergence.2.1}
\end{equation}
where
\begin{equation}
\begin{array}{l}
\eta(\varphi)=\displaystyle\frac \pi 2+
\displaystyle\frac 12\,
\arccos\left[\displaystyle\frac{1-\varphi\,\cot\varphi}
{(1-\varphi\,\cot\varphi)^2+\sin^2\varphi}\right]^{1/4},
\end{array}
\label{convergence.2.2}
\end{equation}
and where $\varphi$ is the solution of the equation
\begin{equation}
\begin{array}{l}
\displaystyle\frac xn=\displaystyle\frac{\sin\varphi}{\varphi}\,
\exp\left(\displaystyle\frac{\varphi}{\sin\varphi}\,\cos\varphi\right).
\end{array}
\label{convergence.2.3}
\end{equation}
For any fixed value of $x$, we are interested in estimating the behavior of 
$\exp(-x)\phi_k(-x)$ for $k\to\infty$. In such limit the solution of 
Eq.~(\ref{convergence.2.3}) tends to $\pi^-$. In the same limit, $\eta(\varphi) \to \pi/2$, so that the whole term $\sin\left[k\left(\pi-\varphi+\displaystyle\frac{\sin^2\varphi}{\varphi}\right)+\eta(\varphi)\right]$ can be replaced by 1. Secondly, in the limit $\varphi\to\pi^-$, we have
\begin{equation}
\begin{array}{l}
\displaystyle\frac 1{\left[\left(\displaystyle\frac \varphi{\sin\varphi}-\cos\varphi\right)^2+\sin^2\varphi\right]^{1/4}}
\approx \left(\displaystyle\frac {\sin\varphi}\varphi\right)^{1/2},
\end{array}
\label{convergence.2.4}
\end{equation}
and
\begin{equation}
\begin{array}{l}
{\exp\left\{k\left[\log\displaystyle\frac{\sin\varphi}{\varphi}
-\displaystyle\frac{\sin\varphi}{\varphi}\cos\varphi\right]\right\}}
\approx \left(\displaystyle\frac {\sin\varphi}\varphi\right)^k\,
\exp\left(k\,\displaystyle\frac{\sin\varphi}{\varphi}\right),
\end{array}
\label{convergence.2.5}
\end{equation}
so that 
\begin{equation}
\begin{array}{l}
\exp(-x)\,\phi_k(-x) \approx k!\sqrt{\displaystyle\frac 2{\pi k}}
\,\left(\displaystyle\frac {\sin\varphi}\varphi\right)^{k+1/2}
\,\exp\left(k\,\displaystyle\frac{\sin\varphi}{\varphi}\right),
\end{array}
\label{convergence.2.6}
\end{equation}
which, once inserted into Eq.~(\ref{convergence.1}),
shows that the asymptotic series displays a factorial divergence.

We now prove that the factorial series in the r.h.s. of 
Eq.~(\ref{convergence.2}) is, for $x>0$, convergent.
This can be done by noting that, for $n\ge 1$,
\begin{equation}
\begin{array}{l}
\displaystyle\frac{k!}{(n)_{k+1}}=
\displaystyle\frac 1{k+1}\,\displaystyle\frac{(k+1)!}{(n)_{k+1}}=
\displaystyle\frac 1{k+1}\,\displaystyle\frac{1\times 2\times \ldots 
(k+1)}{n \times (n +1)\times \ldots (n +k)} \le  
\displaystyle\frac 1{k+1},
\end{array}
\label{generating.7}
\end{equation}
which, by taking into acount the asymptotics of $L^{(-1)}_{k}(x)$ for large $k$, 
namely,
\begin{equation}
\begin{array}{l}
L^{(-1)}_{k}(x)	\exp(-x) \approx \displaystyle\frac 
1{\sqrt\pi}\,\exp(-x/2)\,x^{1/4}\,k^{-3/4}\,\cos\left(2\sqrt{kx}+\displaystyle\frac \pi 4\right),
\end{array}
\label{generating.8}
\end{equation}
leads, for sufficiently high values of $k$, to
\begin{equation}
\begin{array}{l}
\left|\displaystyle\frac{k!\,L^{(-1)}_{k}(x)\,\exp(-x)}{(n)_{k+1}}\right|
\le \displaystyle\frac 1{k+1}\,\left|L^{(-1)}_{k}(x)\,\exp(-x)\right|\\
\\
\le \displaystyle\frac 1{k+1}\, 
\displaystyle\frac{\exp(-x/2)\,\,x^{1/4}}{\sqrt\pi}\,\displaystyle\frac 1{k^{3/4}} 
< \displaystyle\frac{\exp(-x/2)\,\,x^{1/4}}{\sqrt\pi}\,\displaystyle\frac 1{k^{7/4}},
\end{array}
\label{generating.9}
\end{equation}
that proves the absolute convergence of the factorial series.

\subsection{Some remarks for negative values of the ES argument}
\label{negative}

{If $z<0$, the ES in Eq.~(\ref{Euler}) becomes a nonalternating, 
divergent asymptotic power series in $z$, which has been used in 
the literature as a paradigmatic example of a series that \emph{cannot} 
be resummed by the Levin and Weniger transformations~\cite{jentschuraPRD-00}. 
The reason for such inability in the resummation process is strictly 
related to the fact that the negative real axis coincides with the 
branch cut of the exponential integral function $E_1(1/z)$. 
As a matter of fact, the (all positive) single terms of the ES in Eq.~(\ref{Euler})
\emph{cannot} reproduce the imaginary part of the function $E_1(1/z)$ which, for $z<0$, 
is evaluated as\cite{abramowitz}
\begin{equation}
E_1\left(\displaystyle\frac 1z \pm \mathrm{i}0\right)=-\mathrm{Ei}\left(-\displaystyle\frac 1z\right) \mp\mathrm{i}\pi,
\label{negative.1}
\end{equation}
where $\mathrm{Ei}(\cdot)$ denotes the function~\cite{abramowitz}
\begin{equation}
\mathrm{Ei}(x)=\mathcal{P}\,\displaystyle\int_{-\infty}^x\,\displaystyle\frac{\exp(-t)}{t}\,\mathrm{d}t,\;\;\;\;\;\;\;\;\;\;\;\;x>0,
\label{negative.2}
\end{equation}
with the symbol $\mathcal{P}\displaystyle\int_{-\infty}^x\ldots$ denoting the principal value operator,
i.e.,
\begin{equation}
\mathcal{P}\,\displaystyle\int_{-\infty}^x\,\ldots =
\displaystyle\lim_{\epsilon\to 0^+}\,
\left(
\displaystyle\int_{-\infty}^{-\epsilon}\ldots +
\displaystyle\int_{\epsilon}^x\ldots 
\right).
\label{negative.3}
\end{equation}
%
It must be stressed that, for $z<0$, also the asymptotic and the factorial series in Eqs.~(\ref{WT.3}) and~(\ref{generating.5}) become nonalternating, their single terms 
being all positive. For the former this can be proved by noting that the exponential polynomials $\phi_k$ are strictly positive, for any $k$, when their argument is positive, [this can be seen from the recurrence relation in Eq.~(\ref{recurrenceBell.2})]. For the latter, the nonalternating character directly follows from the expansion in Eq.~(\ref{newFormulaStirling.6}). 
On the other hand, from the integral representation in Eq.~(\ref{Euler.1.3.2}), 
it appears that also the converging factor $\varphi_n$ of the ES presents a branch cut 
coincident with the negative real axis $z<0$. Accordingly, we conclude by saying that
for those cases in which the ES cannot be successfully resummed by Levin and Weniger transformations, also the series expansions of $r_n$ in Eqs.~(\ref{WT.3}) and~(\ref{generating.5}) seem to lack their effectiveness in giving a meaningful 
representation of the remainder.  
}

\section{Conclusions}
\label{conclusions}

The understanding of the retrieving action, as well as the development of new types of nonlinear sequence transformations aimed at resumming 
different classed of divergent series requires the large index asymptotics of the corresponding truncation
errors to be investigated. 
A general approach for achieving such task has recently been proposed
in Ref.~\cite{weniger}. In the present paper we showed that, for the (factorially divergent) Euler series, such approach allows the
$n$th-order remainder to be represented via asymptotic and factorial
expansions involving exponential and associated Laguerre polynomials, respectively. The convergence of the above expansions has also been investigated.

\section*{Acknowledgments}

I am indebted to Ernst Joachim Weniger for giving me very useful suggestions. I also thank both reviewers for their remarks
and Turi Maria Spinozzi for his help during the preparation of the manuscript.

\end{document}